\documentclass[a4paper]{jpconf}
\usepackage{graphicx}  
\begin{document}
%
% A.De Santis Definitions
%

% typesetting & referencing
\def\ifm#1{\relax\ifmmode#1\else$#1$\fi}
\newcommand{\refeq}[1]{(\ref{#1})}
\def\fig#1{fig.~\ref{#1}}
\def\Fig#1{Fig.~\ref{#1}}
\def\tab#1{tab.~\ref{#1}}
\def\Tab#1{Tab.~\ref{#1}}
\def\equ#1{eq.~\ref{#1}}
\def\Equ#1{Eq.~\ref{#1}}
\def\etal{{\it et al.}}
\def\red{\color{red}}
\def\blue{\color{blue}}
\def\green{\color{green}}
\newcommand{\bold}[1]{{\bfseries {#1}}}
\def\EG{\emph{e.g.}}

% shortcuts for phys
\def\f{\ifm{\phi}}    
\def\ff{\f--factory}
\def\epm{\ifm{e^+e^-}}
\renewcommand{\to}{\ensuremath{\rightarrow}}
\def\dafne{\ifm{\mbox{DA$\Phi$NE}}}
\def\k2{\ifm{\mbox{KLOE-2}}}
\def\ep{\ifm{e^+}}
\def\el{\ifm{e^-}}
\def\ks{\ifm{\mbox{K}_S}}
\def\kl{\ifm{\mbox{K}_L}}
\def\ko{\ifm{\mbox{K}_0}}
\def\ok{\ifm{\overline{\mbox{K}}_0}}
\def\kp{\ifm{K_+}}
\def\km{\ifm{K_-}}
\def\kplus{\ifm{K^+}}
\def\kmin{\ifm{K^-}}

\def\pio{\ifm{\pi^0}}   
\def\pip{\ifm{\pi^+}}
\def\pim{\ifm{\pi^-}}
\def\neu{\ifm{\nu}}
\def\neubar{\ifm{\bar{\nu}}}
\def\JPC{\ifm{J^{PC}}}
\def\CP{\ifm{CP}}
\def\CPT{\ifm{CPT}}
\def\kltre{\ifm{\mbox{K}_\ell3}}
\def\kmutre{\ifm{\mbox{K}_\mu3}}
\def\ketre{\ifm{\mbox{K}_e3}}
\def\sqrts{\ifm{\sqrt{s}}}
\def\pienu{\ifm{\pi^\pm e^\mp \nu}}
\def\pimunu{\ifm{\pi^\pm \mu^\mp \nu}}
\def\pilnu{\ifm{\pi^\pm l^\mp \nu}}
\def\pipi{\ifm{\pi^+ \pi^-}}

%shotrcuts for math 
\def\ab{\ifm{\sim}}
\newcommand{\ket}[1]{\ifm{|#1\rangle}}
\newcommand{\braket}[2]{\ifm{\langle #1|#2 \rangle}}
\def\order#1,{\ifm{\mathcal{O}(10^{#1})}}
\def\x{\ifm{\times}}

%shortcuts for units
\def\MeV{\ifm{\mbox{MeV}}}
\def\ns{\ifm{\mbox{ns}}}
\def\mV{\ifm{\mbox{mV}}}
\def\GeV{\ifm{\mbox{GeV}}}
\def\mum{\ifm{\mbox{$\mu$m}}}
\def\mm{\ifm{\mbox{mm}}}
\def\mub{\ifm{\mbox{$\mu$b}}}
\def\invfb{\ifm{\mbox{fb}^{-1}}}
\def\invpb{\ifm{\mbox{pb}^{-1}}}
\def\invnb{\ifm{\mbox{nb}^{-1}}}
\def\Amp{\ifm{\mbox{A}}}
\def\Ampsq{\ifm{\mbox{A}^2}}
\def\instlum{\ifm{\mbox{cm}^{-2}\mbox{s}^{-1}}}
\def\l1032{\ifm{10^{32}\mbox{cm}^{-2}\mbox{s}^{-1}}}
\def\degree{\ifm{^\circ}}
\def\MHz{\ifm{\mbox{MHz}}}
\def\Hz{\ifm{\mbox{Hz}}}

%shortcuts for lumi

\def\cbw{\ifm{\Delta_{\mbox{I}}^{\mbox{W}}}}
\def\nmean{\ifm{\mbox{N}_{\mbox{mean}}}}
\def\LTrg{\ifm{\mbox{L}_{\mbox{Trg}}}}
\def\LEst{\ifm{\mbox{L}_{\mbox{Est}}}}
\def\LDel{\ifm{\mbox{L}_{\mbox{Del}}}}
\def\DL{\ifm{\delta\mbox{L}}}

\title{Test of discrete symmetries in transitions with entangled neutral kaons at KLOE-2}

\author{A. De Santis\footnote{on behalf of the KLOE-2 collaboration.}}
\address{Laboratori Nazionali di Frascati, INFN, v. Enrico Fermi, 40, 00044, Frascati (RM).}

\ead{antonio.desantis@lnf.infn.it}

\begin{abstract}
The KLOE-2 experiment at the INFN Laboratori Nazionali di Frascati (LNF) completed its data-taking
at the \epm\ \dafne\ collider, which implements an innovative collision scheme based on a
crab-waist configuration, and achieved the integrated luminosity of more than 5 \invfb.
KLOE-2 represents the continuation of KLOE with an upgraded detector and an extended physics program which includes, among the main topics,  neutral kaon interferometry and test of discrete symmetries . Entangled neutral kaon pairs produced at \dafne\ are a unique tool to test discrete
symmetries and quantum coherence at the utmost sensitivity, strongly motivating the
experimental searches of possible CPT violating effects, which would constitute an unambiguous signal of New Physics. The status of the test of Time reversal and CPT simmetry in \f\to\ks\kl\to$\pi\nu,3\pio,(2\pi)$ decays with KLOE and KLOE-2 data will be discussed.
\end{abstract}

\section{Introduction}

\dafne , the Frascati \ff, is an \epm\ collider working at a center of mass energy of
\sqrts\ab1020\MeV\cite{dafne},  corresponding to the peak of the \f\
resonance.   The KLOE experiment at \dafne\ completed its first data taking campaign in March 2006
with a total integrated luminosity of \ab2.5 \invfb, corresponding to a production of
\ab7.5$\times 10^9$ \f-mesons and \ab2.5$\times 10^9$ \ko\ok\ pairs.
After the KLOE run, \dafne\ has been upgraded implementing an innovative collision scheme
based on a crab-waist configuration \cite{dafne2}.  The \k2\ experiment \cite{k2physprog},
aiming to extend the physics program of its predecessor, completed the data-taking in March
2018 at the upgraded \dafne\ with an improved detector. The total integrated luminosity
collected was \ab5.5 \invfb, as originally planned.
The \k2\ physics program includes neutral kaon interferometry and tests of discrete symmetries and
quantum mechanics.
\par
The properties of the neutral kaon system are directly related to the CP, T and CPT
symmetries and provide the potential of performing very precise tests and to search for violation
effects. The quantum entanglement of neutral kaons produced by the \f\ decay, allows for a large number of quantum interferometry studies.
The KLOE experiment, is the only experiment at \ff's, so has the unique possibility to study
the entangled neutral kaon pairs and to give a large contribution to the knowledge of kaon physics
and related discrete symmetries violation.

\section{The KLOE-2 experiment}

The original KLOE detector consists of a large cylindrical drift chamber (DC) \cite{Adinolfi:2002uk},
which provides excellent momentum and vertex reconstruction accuracy for charged particles. DC is surrounded
by a lead-scintillating fiber electromagnetic calorimeter (EMC) \cite{Adinolfi:2002jk}.
The energy deposits of charged and neutral particles in the calorimeter are measured with very
good time resolution, allowing particle identification with time-of-flight (TOF) techniques.
A superconducting coil around the EMC provides a 0.52 T axial field. 
\par
The upgrade of the KLOE detector was based on: i) an inner tracker (IT) made of 
cylindrical GEM for the improvement of tracking and decay vertex resolution close
to the interaction point (IP)\cite{Balla:2014kqa}, ii) two \epm\ tagging system for the $\gamma\gamma$
physics at low and high lepton energy regimes \cite{Babusci:2009sg,Archilli:2010zza},
a pair of crystal calorimeters inside the innermost part of the detector, close to IP,
to increase the photon acceptance down to 8\degree\ \cite{Cordelli:2013mka}, iii) a pair of
scintillator/absorber detectors surrounding the beam pipe region 
\cite{Cordelli:2009xb} to improve acceptance and efficiency for photons and pions coming
from neutral kaon decays.
\par
Kaon physics is typically studied by tagging one the two kaons with a special decay/interaction of the other. This is the case of \ks\ tagged by the \kl\ interaction in the EMC calorimeter or the \kl\ tagged  by the \ks\ decay near the IP.
Nevertheless a different approach to the kaon physics is possible at
\ff, based on the observation of the time evolution of the system correlation.
The quantum mechanics description of the \f\to\ks\kl\ decay implies an anti-correlated
initial state that evolves in time preserving this characteristics. This
feature has been already exploited to perform several tests and measurements on the
kaon system \cite{Ambrosino:2006vr,Babusci:2013gda}.
Kaon correlation could be also used to tag CP or Flavor eigenstate during the time
evolution of the initial state \cite{Bernabeu:2012nu}, as discussed in the next section. 

%\section{Time dependent tagging}
\section{Discrete symmetry tests in kaon transition amplitudes}

% riscrivere
The quantum correlation between the two neutral kaons allows the time-tagging of the initial state
of one of the two by using the decay of the other as shown in \fig{fig:evolution} (left).
In the sketch the $f_1^\alpha$ is the tagging decay observed at the time $t_1$ that implies a well
defined corresponding state for the undecayed kaon ($\bar{\mbox{K}}_\alpha$) at the same time.  
Similarly the ``tagged'' kaon will be observed to decay in the final state $f_2^\beta$ at the time
$t_2$ as  sketched in \fig{fig:evolution} (right). This decay will reveal the state of the second kaon
as ${\mbox{K}}_\beta$, so the  transition amplitudes between $\bar{\mbox{K}}_\alpha$ and ${\mbox{K}}_\beta$
could be derived from the observed time evolution between $t_1$ and $t_2$.

\begin{figure}[h!]
    \centering
    \includegraphics[width=0.43\textwidth]{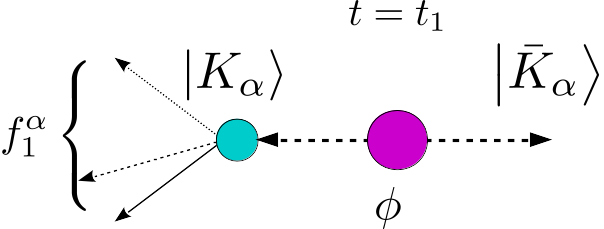}
    \includegraphics[width=0.55\textwidth]{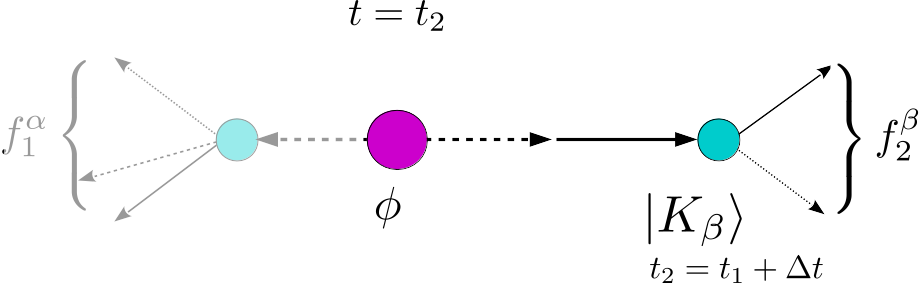}
    \caption{Time evolution of the kaon system at the time of the first decay $t_1$ 
      in the final state $f_1^\alpha$ (left) and at the time $t_2$ of the second decay $f_2^\beta$.
      The entanglement between the two kaons allows to identify the state of the undecayed kaon
      at $t_1$. The observation of the decay at $t_2$ allows to study the transition amplitudes
      between the initial and final state.}
      \label{fig:evolution}
\end{figure}

Different transition amplitudes can be studied with the corresponding choice of the kaon decay
pairs. In the table \ref{tab:transitions} all possible transition amplitudes between flavor and \CP\
eigenstates are related via the corresponding discrete symmetries. A comparison of the rates
of neutral mesons transitions between their flavor and \CP\ eigenstates allows for a model
independent test of the T and CPT symmetries. Similar test has been performed already in the case
of neutral B mesons obtaining the first direct evidence of T violation \cite{Lees:2012uka}.

\begin{center}
  \begin{table}[h]
    \label{tab:transitions}
    \caption{Transition amplitudes connected via discrete symmetries. \ko\ok\ and
      \kp\km\ are Flavor and CP eigenstates, respectively.}
    \centering     
    \begin{tabular}{c|c|c|c}
      \br
      Reference & T-conjug. & CP-conjug. & CPT conjug. \\      
      \br
      \ko\to\kp & \kp\to\ko & \ok\to\kp & \kp\to\ok \\
      \ko\to\km & \km\to\ko & \ok\to\km & \km\to\ok \\
      \ok\to\kp & \kp\to\ok & \ko\to\kp & \kp\to\ko \\
      \ok\to\km & \km\to\ok & \ko\to\km & \km\to\ko \\
      \br
    \end{tabular}
  \end{table}
\end{center}

As stated previously the kaon states along the time evolution are identified by using the decay
channel. The flavor eigenstates are identified by using the semileptonic decays \ko\to\pim\ep\neu\ and
\ok\to\pip\el\neubar\ because the charge of the lepton emitted in the decay is connected with the sign of the intermediate W boson responsible for the decay at tree level.
The \CP\ eigenstate instead are tagged by using the fully hadronic decay mode in two (\kp\to\pip\pim)
or three pions (\km\to3\pio). The observables related to T and CPT violation are defined as:

\begin{eqnarray}
  R_2(\Delta t) & = & \frac{P(\ko(0)\to K_-(\Delta t))}{P(K_-(0)\to \ko(\Delta t))} 
  \sim  \frac{I(l^-,3\pi^0;\Delta t)}{I(\pi\pi,l^+\Delta t)} \label{eq:tvr2} \\ 
  R_4(\Delta t) & = & \frac{P(\ok(0)\to K_-(\Delta t))}{P(K_-(0)\to \ok(\Delta t))}
  \sim \frac{I(l^+,3\pi^0;\Delta t)}{I(\pi\pi,l^-\Delta t)}  \label{eq:tvr4}  \\ 
  R_2^{CPT}(\Delta t) & = & \frac{P(\ko(0)\to K_-(\Delta t))}{P(K_-(0)\to \ok(\Delta t))} 
  \sim \frac{I(l^-,3\pi^0;\Delta t)}{I(\pi\pi,l^-\Delta t)} \label{eq:cptvr2} \\ 
  R_4^{CPT}(\Delta t) & = & \frac{P(\ok(0)\to K_-(\Delta t))}{P(K_-(0)\to \ko(\Delta t)}
  \sim  \frac{I(l^+,3\pi^0;\Delta t)}{I(\pi\pi,l^+\Delta t)} \label{eq:cptvr4}     
\end{eqnarray}

where $I(f_1 ,f_2 ;\Delta t)$ denotes the number of recorded events characterized by a time-
ordered pair of kaon decays $f_1$ and $f_2$ separate by an interval of proper kaon decay
times $\Delta t$. A deviation of the asymptotic level of these ratios from unity for
large transition times would be a T or CPT violation manifestation. Preliminary results on
the time reversal symmetry tests are reported in the \Fig{fig:r2r4_tv} where the distribution
relative to \equ{eq:tvr2} and \equ{eq:tvr4} are shown.
\par
A more robust test on the CPT symmetry violation is shown in \Fig{fig:r2r4_cpt} where the
double ratio $R_2^{CPT}(\Delta t)/R_4^{CPT}(\Delta t)$  is shown. The double ratio is built from
Eq.~\ref{eq:cptvr2} and \ref{eq:cptvr4} and allows to cancel many systematic effect related
to tracking, particles identification and decay vertex reconstruction.

\begin{figure}[h!]
    \centering
    \includegraphics[width=0.75\textwidth]{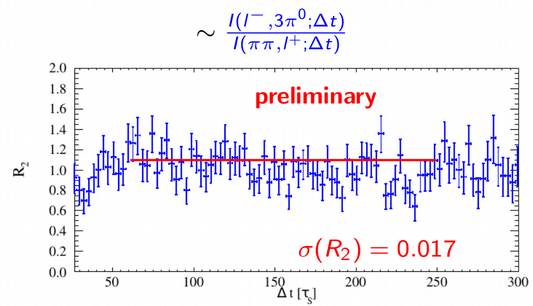}
    \includegraphics[width=0.75\textwidth]{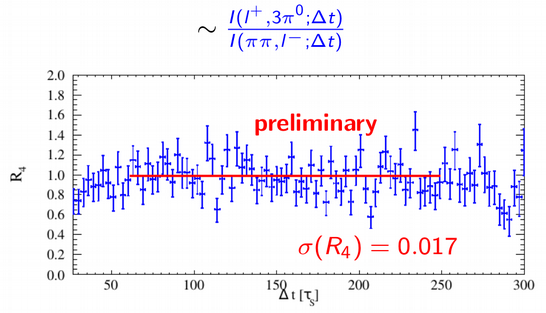}
      \caption{Preliminary distribution of the T-violation sensitive ratios of
        neutral kaon double decay rates as a function of kaon decay times
        difference ($\Delta t$) as in Eq.~\ref{eq:tvr2} and \Equ{eq:tvr4}.
        The statistical uncertainty of the asymptotic level of this observable
        for $\Delta t \gg \tau_S$ (red line) is relative to the KLOE data-set
        (2 \invfb) only.}
      \label{fig:r2r4_tv}
\end{figure}

\begin{figure}[h!]
    \centering
      \includegraphics[width=0.75\textwidth]{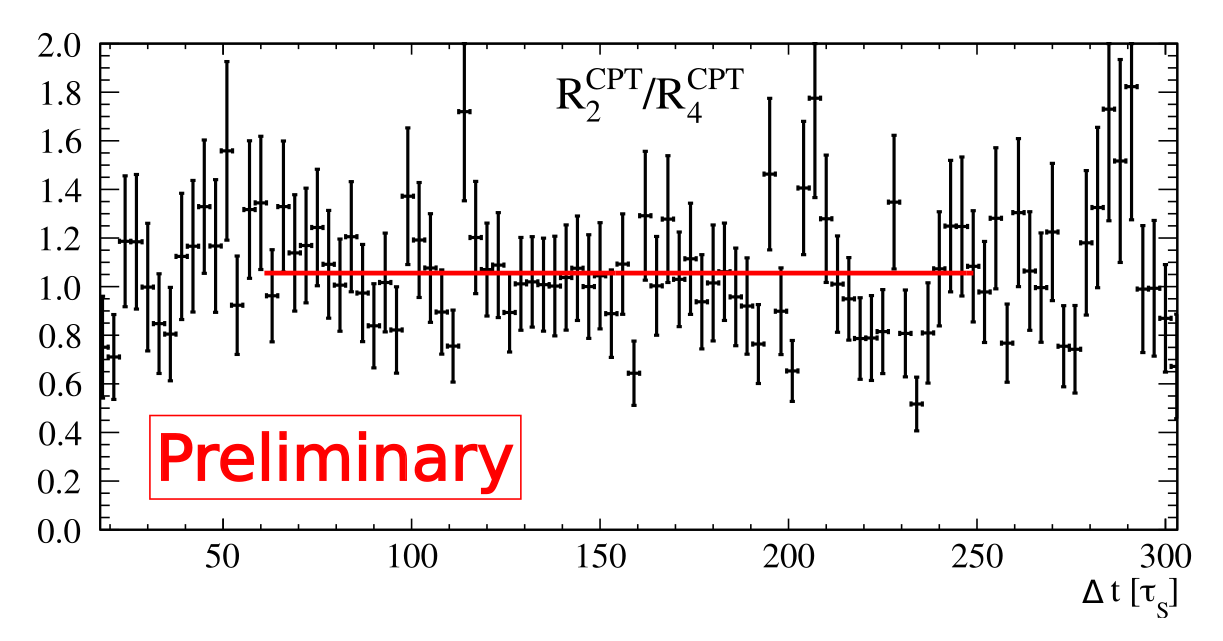}
      \caption{Preliminary distribution of the double ratio $R_2^{CPT}(\Delta t)/R_4^{CPT}(\Delta t)$
        of neutral kaon double decay rates as a function of kaon decay times
        difference ($\Delta t$) defined in Eq.~\ref{eq:cptvr2} and \Equ{eq:cptvr4}.
        This distribution is sensitive to CPT-violation for large $\Delta t$
        The statistical uncertainty of the asymptotic level of this observable
        for $\Delta t \gg \tau_S$ (red line) amounts to 0.011 with the KLOE data-set (2 \invfb).}
      \label{fig:r2r4_cpt}
\end{figure}

\section{Conclusions}

The data-taking finished in 2018 allows the \k2\ collaboration to access a unique
data-set of order of 8 \invfb\ of \f\ decays. Among all the possible studies, the kaon
quantum correlation, already studied with the old KLOE data-set, will be further
explored as the presented study shows. The expected precision of $10^{-3}$ on the
double ration will be achievable when the full data-set will be analyzed and all
the systematics will be carefully taken into account.

\section*{Acknowledgments}
We warmly thank our former KLOE colleagues for the access to the data collected during the KLOE data taking campaign.
We thank the DA$\Phi$NE team for their efforts in maintaining low background running conditions and their collaboration during all data taking. We want to thank our technical staff: 
G.F. Fortugno and F. Sborzacchi for their dedication in ensuring efficient operation of the KLOE computing facilities; 
M. Anelli for his continuous attention to the gas system and detector safety; 
A. Balla, M. Gatta, G. Corradi and G. Papalino for electronics maintenance; 
C. Piscitelli for his help during major maintenance periods. 
This work was supported in part 
by the Polish National Science Centre through the Grants No.\
2013/11/B/ST2/04245,
2014/14/E/ST2/00262,
2014/12/S/ST2/00459,
2016/21/N/ST2/01727,
2016/23/N/ST2/01293,
2017/26/M/ST2/00697.

\end{document}